# Deep Lead Optimization: Leveraging Generative AI for Structural Modification

Odin Zhang[1,#], Haitao Lin[2,#], Hui Zhang[1], Huifeng Zhao[1], Yufei Huang[2], Yuansheng Huang[1], Dejun Jiang[1], Chang-yu Hsieh[1,*], Peichen Pan[1,*], Tingjun Hou[1,*]

[1]College of Pharmaceutical Sciences, Zhejiang University, Hangzhou 310058, Zhejiang, China
[2]AI Lab, Research Center for Industries of the Future, Westlake University, Hangzhou, China
[#]Equal Contributions

## Corresponding authors

**Tingjun Hou**
E-mail: tingjunhou@zju.edu.cn
**Chang-Yu Hsieh**
E-mail: kimhsieh@zju.edu.cn
**Peichen Pan**
E-mail: panpeichen@zju.edu.cn

# Abstract

The idea of using deep-learning-based molecular generation to accelerate discovery of drug candidates has attracted extraordinary attention, and many deep generative models have been developed for automated drug design, termed molecular generation. In general, molecular generation encompasses two main strategies: *de novo* design, which generates novel molecular structures from scratch, and lead optimization, which refines existing molecules into drug candidates. Among them, lead optimization plays an important role in real-world drug design. For example, it can enable the development of me-better drugs that are chemically distinct yet more effective than the original drugs. It can also facilitate fragment-based drug design, transforming virtual-screened small ligands with low affinity into first-in-class medicines. Despite its importance, automated lead optimization remains underexplored compared to the well-established *de novo* generative models, due to its reliance on complex biological and chemical knowledge. To bridge this gap, we conduct a systematic review of traditional computational methods for lead optimization, organizing these strategies into four principal sub-tasks with defined inputs and outputs. This review delves into the basic concepts, goals, conventional CADD (computer-aided drug design) techniques, and recent advancements in AIDD (AI-aided drug discovery). Additionally, we introduce a unified perspective based on constrained subgraph generation to harmonize the methodologies of *de novo* design and lead optimization. Through this lens, these two areas are seen as complementary, with each capable of enhancing the other: de novo design can incorporate strategies from lead optimization to address the challenge of generating hard-to-synthesize molecules; inversely, lead optimization can benefit from the innovations in de novo design by approaching it as a task of generating molecules conditioned on certain substructures. In conclusion, we spotlight the challenges and promising directions for future research, offering valuable insights and resources for both machine learning and chemistry communities.

# Introduction

It is well acknowledged that drug development is an extremely labor-intensive and time-consuming process. Estimates suggest that the economic cost of developing a drug is about $1 billion and over a period lasting more than ten years. On average, only five out of 5,000 candidate compounds can successfully enter clinical trials, with just one receiving regulatory approval to become a marketable drug[1]. In general, a clinical drug has to go through three intertwined stages: hit[2], lead[3], and candidate[4]. Finally, a properly optimized compound, which may look substantially distinct from the starting compound, has a chance to proceed to clinical research and trial. At the early stages of drug discovery, computer-aided drug design (CADD) can efficiently identify hits and leads with high activity and favorable drug-like properties and significantly expedite the drug development process. In recent years, with the expansion of databases and the rise of artificial intelligence (AI), a new driving force in CADD, AI-aided drug discovery (AIDD), has been utilized to push the boundaries of CADD[5-7]. In both CADD and AIDD, two complementary approaches, virtual screening and molecular generation, can be leveraged to identify or design inhibitors. Virtual screening involves evaluating and ranking a library of compounds to select the most promising candidates for further bioassay testing. On the other hand, molecular generation ventures beyond the limitations of existing compound libraries to design new chemical entities, thereby expanding the chemical space with novel possibilities, as illustrated in **Figure 1**. Compared to virtual screening, molecular generation shows promise in designing novel drugs, and has been successfully applied in several real-world drug discovery campaigns[8,9].

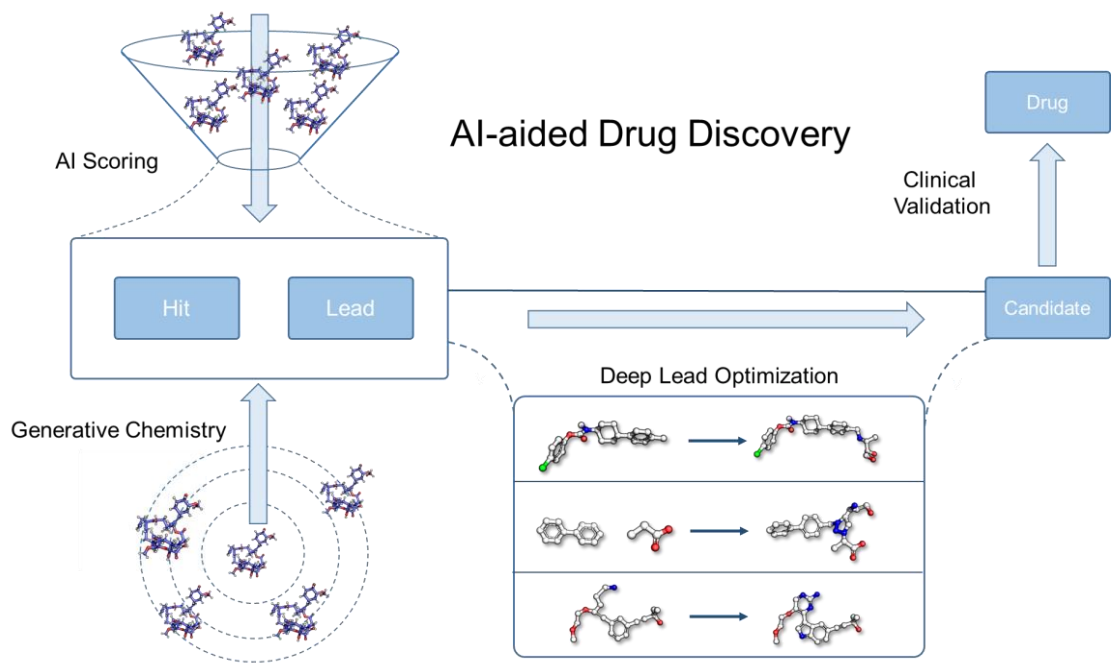

**Figure 1.** The role of deep lead optimization in the pipeline of AI-aided Drug Discovery.

Molecular generation bifurcates two streams: *de novo* design and lead optimization. De novo design involves the creation of entirely new molecular structures without relying on pre-existing templates, aiming to explore uncharted areas of the chemical space for potential therapeutic agents. On the other side, lead optimization focuses on refining and improving existing lead compounds to enhance their efficacy, selectivity, pharmacokinetics, and safety profiles. This process iteratively modifies chemical structures, using insights from biological testing and computational predictions to guide modifications, with the goal of converting leads into promising drug candidates. Given the complexity and targeted nature of lead optimization, it can be chemically classified into four pivotal sub-tasks. These include scaffold hopping[10], which replaces core structures while retaining functional activity; linker design[11], optimizing connections between molecular fragments; fragment replacement[12], substituting parts of the molecule to enhance binding; and side-chain decoration[13], modifying mutiple side chains while keeping privileged scaffolds. A brief overview of these tasks is presented in **Figure 2**. Many pharmaceutical companies and biotechs rely heavily on lead optimization to design me-too or me-better drugs which are novel enough to overcome patent protection. For instance, atorvastatin[14], atorvastatin, the fifth marketed cholesterol drug, is superior to the first-in-class lovastatin; and levofloxacin, an antibacterial drug, is superior to the first-in-class norfloxacin[15]. Consequently, the development of reliable and powerful lead optimization tools is beneficial for both academia and industry.

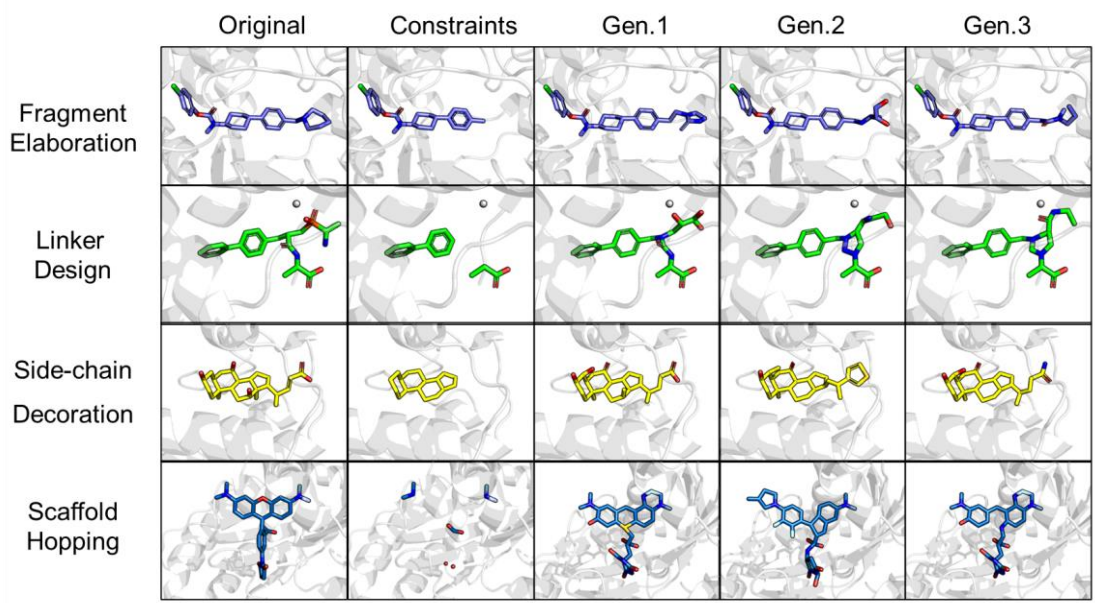

**Figure 2.** Illustration of four subtasks of lead optimization. Each column refers to original compounds, constrained substructures, and generated examples.

In recent years, there has been a growing interest in generative AI, particularly in formulating molecular design as a graph generation problem. For instance, in de novo design, models such as CVAE[16], which utilizes the Variational Autoencoder[17] architecture, MoFlow[6], based on the Flow[18] model, and MolGAN[19], leveraging the GAN[20] architecture, exemplify the advancements in this area. However, lead optimization, despite its critical role in drug design, remains relatively underexplored. Only in the last few years have some lead optimization algorithms based on graphs and Transformers begun to emerge, such as DeLinker[21], DeepFrag[22], DeepHop[23]. Yet, these efforts typically address specific sub-tasks, highlighting the need for a comprehensive review and outlook in this field. In this paper, we aim to fill this gap by introducing four lead optimization scenarios and summarizing the goals, CADD approaches, and

AIDD models developed for each task. In the initial section, we present the concept of lead optimization as a graph-constrained generation problem, drawing parallels between de novo design and lead optimization through a "sketch and fill" analogy. We then provide an overview of the goal and development of four distinct lead optimization strategies, as well as their definition and application within computational models. Subsequently, we offer a comprehensive summary of 32 existing deep-learning methods, which correspond to our categorization of lead optimization techniques. Finally, we envision the future trajectory of this field and urge the machine-learning community to actively contribute toward real-world applications that are of significant interest to pharmacologists.

## Intertwined de novo Generation and Lead Optimization: a Metaphor

From a modeling perspective, a succinct summarization of lead optimization could be "by leveraging existing structures to guide the design of new molecules with enhanced properties". Side chain decoration and fragment replacement, for instance, involve removing the original side chains or the intended replaced fragment from the whole molecule and generating corresponding structures using computational methods. Similarly, linker design and scaffold hopping can be achieved by complementing the bridging of separate molecular fragments.

To harmonize *de novo* generation and lead optimization, we formulate the two concepts in the language of deep generative modeling for better comparison. Mathematically, *de novo* generation learns $P_S(G)$, i.e., the distributions of molecular graphs $G$ over a particular chemical space $S$, while lead optimization aims to model the conditional distribution $P_S(G_\mu|G_{\setminus\mu})$, learning the distributions of remaining structures $G_\mu$ given a partial structure $G_{\setminus\mu}$ over a particular chemical space. One advantage of lead optimization is that it is easier to model the conditional distribution $P_S(G_\mu|G_{\setminus\mu})$ compared to the marginal distribution $P_S(G)$, as the partial structure $G_{\setminus\mu}$ provides a reliable and focused starting point for generating new structures. An analogy can be drawn by considering the difficulty of drawing a portrait from scratch versus filling in detail given a rough sketch of an object, as illustrated in Figure 2. If you are a beginner in painting and this is your first time attempting to paint the Mona Lisa from scratch, you are likely to end up with cartoon-like sketches. However, if you already have a draft and want to add some details to it, the task would become much easier. This is a direct intuition that explains why molecules generated by lead optimization models are typically more valid than those from *de novo* design, and why many medicinal chemists prefer designing molecules based on lead compounds.

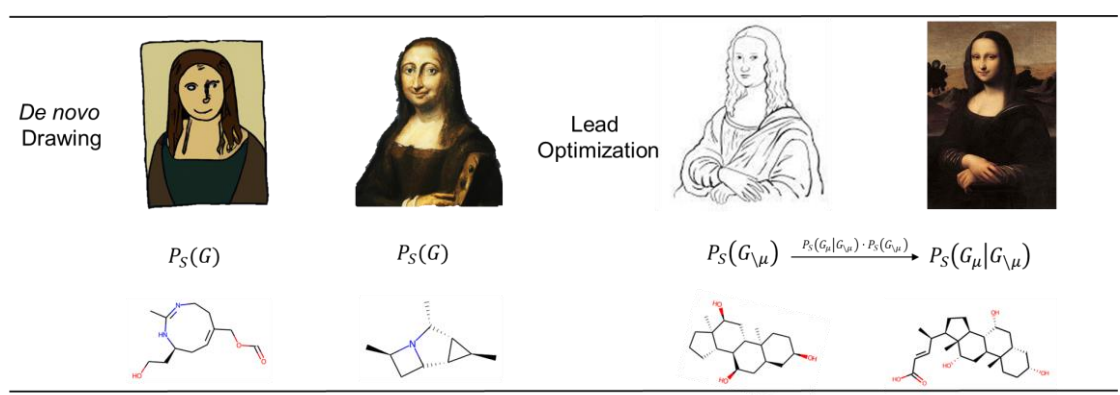

**Figure 3.** The metaphor of de novo generation versus lead optimization.

The distribution $P_S(G)$ can be decomposed into $P_S(G_\mu|G_{\setminus\mu}) \cdot P_S(G_{\setminus\mu})$, implying that for comprehensive coverage of chemical space using lead optimization models, the prior distribution $P_S(G_{\setminus\mu})$ should be highly expressive, or the variety of parent skeletons should be notably broad. To elaborate further, we can imagine a picture where each partial structure $G_{\setminus\mu}$ represents a point within the chemical space, and generating a complete molecule $G$ is to sample from the conditional distribution $P_S(G_\mu|G_{\setminus\mu})$, centered around the point $G_{\setminus\mu}$. For effective coverage of chemical space, crucial for most molecular generation models, the distribution of $G_{\setminus\mu}$ should uniformly cover the chemical space. The established analysis[24] has shown that the number of core structures $G_{\setminus\mu}$ of molecules generally increases linearly with compound library size. Consequently, for direct use of lead optimization models in exploring chemical space, researchers should first examine the fragment library composed by $G_{\setminus \mu}$. However, in the context of lead optimization, similar structures tend to yield similar pharmaceutic effects. Therefore, we need several $G_{\setminus \mu}$ rather than a comprehensive fragment library to reflect the active areas targeting specific protein targets within the entire chemical space.

In summary, we introduce a perspective of constrained graph generation for lead optimization and provide a vivid metaphor to illustrate its significance in real-world applications. Within this framework, numerous advancements in generative AI models can be employed to address the lead optimization problem, such as image inpainting using latent diffusion[25] or image completion with GAN architecture[26].

## Molecular Decomposition

Training a deep learning model requires a substantially large dataset. However, extracting structures of lead compounds before and after optimization from pharmacological articles is exceedingly laborious, resulting in a lack of dedicated datasets for lead optimization. Consequently, it was necessary to leverage computational methods to define lead optimization training data. Specifically, to simulate the process of lead optimization, each data entry should contain a pair of molecules, corresponding to the molecules before and after optimization. In the following section, we will thoroughly discuss how to construct appropriate molecular pairs (the data entry) for lead optimization tasks, while **Figure 4** visualizes these molecular decomposition methods.

### Scaffold and Sidechain Decomposition

Molecules are generally composed of ring structures, linkers and side chain groups/atoms. Ring structures are highly prevalent in molecular drugs, as evidenced by a study[27], in PJB pharmaprojects, 96 percent of chemical entities contain ring structures and 56 percent of the molecular weight is contributed by these ring components. Ring structures are also considered to be the primary synthetic units in organic chemistry and play a significant role in determining the shape, electron distribution and biological activity of a compound[28]. Therefore, the ring structure is regarded as a fundamental unit of a molecular backbone.

The scaffold of a molecule may be defined in various ways, but the most commonly adopted definition is proposed by Bemis and Murcko[29]. The BM scaffold is described as structures that remain after removing all the terminal acyclic side chains, as exemplified in **Figure 4A**. By neglecting the types of atoms and bonds, a more abstract cyclic skeleton (CSK) can be obtained. BM and CSK scaffolds are favorable when analyzing the diversity of compound libraries. The original BM scaffold paper[29] found that about half of all drugs available at that time could be represented by a mere 32 scaffolds, which further supports that scaffolds are expressive enough in outlining the complex chemical space. While BM scaffold is the prominent choice in medicinal chemistry, its partitioning, where the core structure contains all rings while side chains lack cyclic structures, is sometimes considered too crude. To overcome this issue, more elaborate decomposition methods have been developed, e.g., HierS[24], SCONP[30] followed by Sacffold Tree[31], CSE[32], Scaffold Netwrok[33], and Scaffold Hunter[34] which is still available online (as of April, 2023). These methods produce slightly different molecular scaffolds, as shown in **Figure 4B-D**, which are more nuanced decomposition practices and are favorable to cluster, analyze, and understand intrinsic properties of molecules at different levels.

In summary, all these methods to define scaffolds share a central idea, i.e., further disentangling of the ring structure in the BM scaffold. Assuming that the BM scaffold has five-ring structures (**Figure 4A**), further scaffold extraction breaks the single bond connected to the ring to split it into a four-ring structure (**Figure 4B**) and a two-ring structure (**Figure 4D**), in which case the BM scaffold is called the super-scaffold, while the skeleton of the minimal ring structure is called the basis, and the scaffold between the super and basis is called the intermediate. Generally speaking, the granularity of the super-scaffold is too large, and that of the basis scaffold is too small to adequately characterize the chemical space of drug-like molecules. Thus, they are not often used in chemical space exploration and compound clustering analysis[24]. Additionally, different approaches treat fused ring structures differently. For example, the HierS treats the fused ring directly as a whole, while the ScaffoldTree further dissects them, leading to a different structure of the scaffold tree or scaffold hierarchy network.

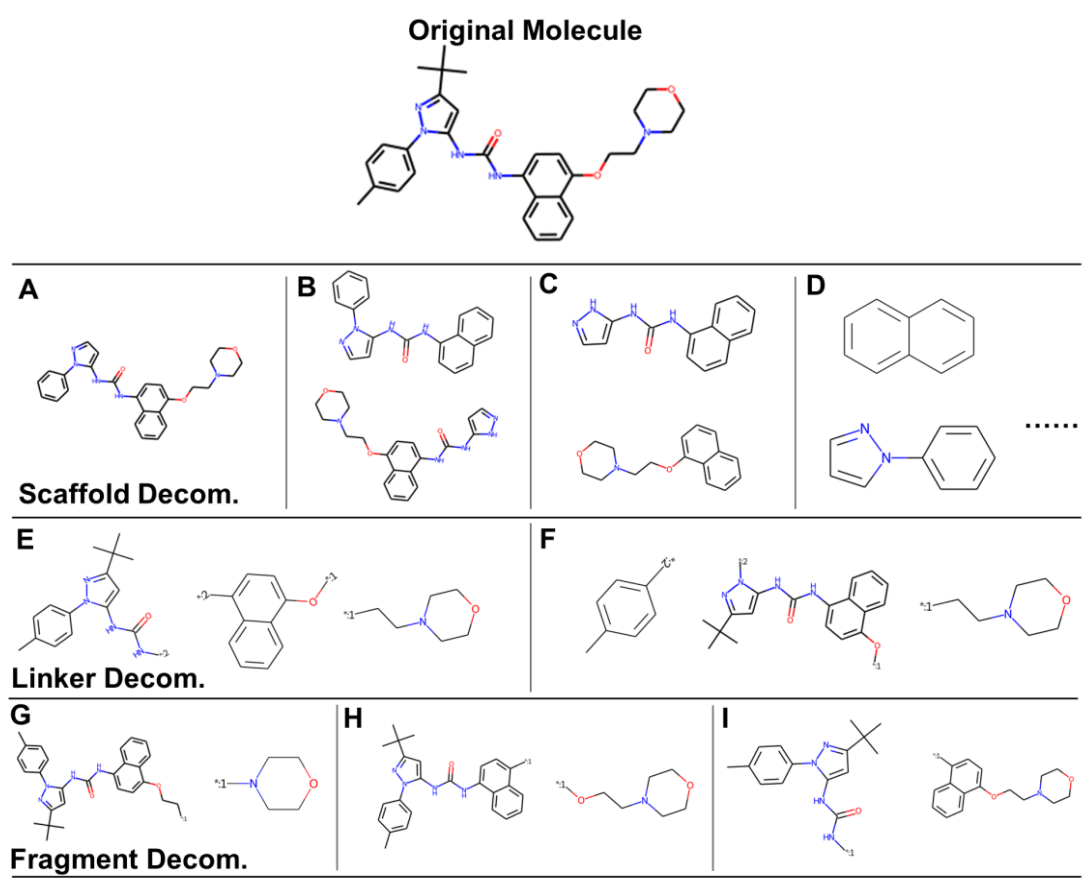

**Figure 4.** Molecular decomposition methods for lead optimization training pairs. A) refers to the B-M scaffold; B-D) refers to more sophisticated scaffold decompositions.

In both scaffold hopping and side-chain decoration tasks, scaffolds are usually defined as the BM Scaffold, while side-chains are often decomposed to non-cyclic terminal chains. Thus, training pairs for Scaffold hopping can be curated to include various 3D similar molecules with distinct 2D structures (each molecule having one BM scaffold). On the other hand, for side-chain decoration, the dataset comprises different molecules sharing an identical scaffold.

**Fragment and Linker Breaking**

In the fields of combinatorial chemistry[35, 36], retrosynthesis route planning[37, 38], and fragment-based drug design[39, 40], molecules are often broken down into basic building blocks known as fragments. Two representative strategies to chemically break up molecules are RECAP[41] and BRICS[42]. RECAP defines 11 types of breakable bonds relevant to chemical reactions, which can be used to dissect complex molecules into their basic constituents. BRICS then goes further and expands the 11 breakable bonds to 16 based on more refined rules (e.g., accounting for substructures in the vicinity of the broken bonds). Another fragmentation method, eMolFrag[43] normalizes fragments and linkers among split structures and has been validated on adenosine receptors, demonstrating that its cleaved units can assemble into active molecules.

Except for fragmenting molecules in a chemical synthesis style, among the deep lead optimization methods, another choice is the use of matched molecular pairs (MMP) method[44] to derive fragments and linkers. MMP was initially proposed to assess the impact of substituent groups on biological activity by focusing on the dissimilarity between molecular pairs. To accomplish this, MMP obtain molecular fragments by defining a set of acyclic single bonds in the pre-cut molecule as $\{b_i\}_{i=1}^{N}$, and randomly selecting one for partitioning the molecule into two fragments. More generally, breaking two bonds in a molecule results in the formation of three fragments, while breaking three bonds produces four fragments, and so on. The linkers derived from MMP constitute the middle portion connecting the two fragments, facilitating the standardization of data for model training. As a result, most of the current linker generation models[21, 45, 46] employ this approach to create lead optimization pairs. However, it is important to note that these linker generation models can only design linkers that are positioned between two fragments, and cannot generate linkers between three or more pieces.

## Lead Optimization: From Goals to Tools

**Table 1** presents the classifications of deep learning-based lead optimization models. The first region of the table is categorized based on the task classification, while the second region is categorized based on whether the model generates the 3D conformation of the molecule and whether the protein structure is given as a conditional constraint.

Table 1. Classifications of deep lead optimization.

| Methods | Linker Design | Scaffold Hopping | Side-chain decoration | Fragment Replacement | Binding Pose (3D) | Pocket-aware |
|---|---|---|---|---|---|---|
| DeLinker | √ | | | | | |
| SyntaLinker | √ | | | | | |
| DRLinker | √ | | | | | |
| Link-INVNET | √ | | | | | |
| ShapeLinker | | | | | | |
| DiffLinker | √ | | | | √ | √ |
| 3DLinker | √ | | | | √ | |
| LinkerNet | √ | | | | √ | √ |
| GraphGMVAE | | √ | | | | |
| DeepHop | | √ | | | | |
| ScaffoldGVAE | | √ | | | | |
| DiffHopp | | √ | | | √ | √ |
| GNNGAC | | | √ | √ | | |
| DeepScaffold | | | √ | √ | | |
| GraphScaffold | | | √ | √ | | |
| MoLeR | | | √ | √ | | |
| 3DScaffold | | | √ | √ | √ | |
| 3DScaffold-RL | | | √ | √ | √ | √ |
| MolGPT | | | √ | | | |
| SmilesScaffold | | | √ | √ | | |
| SCMG | | | √ | √ | | |
| SAMOA | √ | | √ | √ | | √ |
| LibINVENT | | | √ | √ | √ | √ |
| DiffDec | | | √ | √ | √ | √ |
| DeepFrag | | | | √ | | |
| STRIFE | | | | √ | | |
| Transformer-R | | | | √ | | |
| DEVELOP | √ | | | √ | | |
| D3FG | | | | √ | √ | √ |
| SAFE | √ | √ | √ | √ | | |
| DrugEX-v3 | √ | √ | √ | √ | √ | √ |
| REINVENT4 | √ | √ | √ | √ | √ | √ |
| Delete | √ | √ | √ | √ | √ | √ |

**Linker Design**

Linker design is not only one part of the lead optimization problem but also one of the difficulties in fragment-based drug discovery (FBDD)[11, 47], which first screens for lead fragments with low molecular weight (below 260 Da). While lead hit compounds typically exhibit weak binding affinity (~mM), they strongly interact with key residues within protein pockets through polar interactions[48]. Therefore, optimization of these lead fragments lies in the design of suitable linkers to bridge them together, and it is expected that the binding pattern of the lead fragments to the pocket could be retained, allowing the well-known superadditivity of the linked lead, which can be explained by the following content:

$$\Delta G_{A-B} = \Delta G_A + \Delta G_B + \Delta G$$

where $\Delta G_{A-B}$ is the change of free energy of the linked molecule, $\Delta G_A$ and $\Delta G_B$ is that of fragments A and B, while $\Delta G$ is the additional free energy change contributing to the superadditivity. For example, in the development of inhibitors targeting the alpha-catalytic site of tyrosine kinase 2[49, 50], the lead fragments exhibited $K_d$ values of 270 μM and >500 μM, respectively. After being connected by an elaborate linker, the activity of the optimized lead increased to 0.320 μM, representing a 1,000-fold increasement. This significant enhancement drew the attention of medicinal chemists, leading to the evolution of two classical linking methods, library search[51] and FMO calculation[52].

The first deep learning model that focuses on linker design is DeLinker[21], which uses a VAE architecture[17] as shown in **Figure 5A**. In this framework, assuming that the training molecule contains $n$ atoms, including $v$ fragment atoms and $m$ linker atoms, Delinker encodes the fragment structure combined with the entire molecule to latent space, and then decodes it to recover the original molecular structure. The linker generation is done by encoding the fragment structure into latent space and then decoding it in combination with a noise vector of $v + m$ dimensions to produce a molecule with a $m$ atoms linker. The SyntaLinker[53] approach as shown in **Figure 5B**, which was developed later, uses the language model Transformer[54] for Linker design by generating SMILES. With encoding fragmented smiles and conditionally controlled markers (shortest linkage paths and other pharmacophore constraints), SyntaLinker can decode complete molecules that meet certain constraints and can also be extended to target-specific application scenarios. SyntaLinker outperforms DeLinker in terms of validity, rationality, and recovery probability owing to the powerful learning capability of Transformer backbone[54]. Afterward, SyntaLinker-Hybrid[55] was introduced, which uses fragment hybridization to refine the fragment-based drug design approach using the SyntaLikner, but the model itself was not fundamentally improved. DEVELOP[56] is an upgraded version of DeLinker, which merges linker design and fragment elaboration tasks into a single model. It incorporates 3D pharmacophore constraints into molecular generation using CNN[57] as the feature extractor, enabling ligand-based drug design. Another notable model, 3DLinker[58], refinedly considers linker geometry, exploiting equivariant neural networks to obtain the coordinate of the linked atom at each iterative generation step. Compared to models like DeLinker, which can only generate two-dimensional structures, 3DLinker performs better on various geometry-aware metrics, enhancing the ability to engineer linkers with dominant geometries.

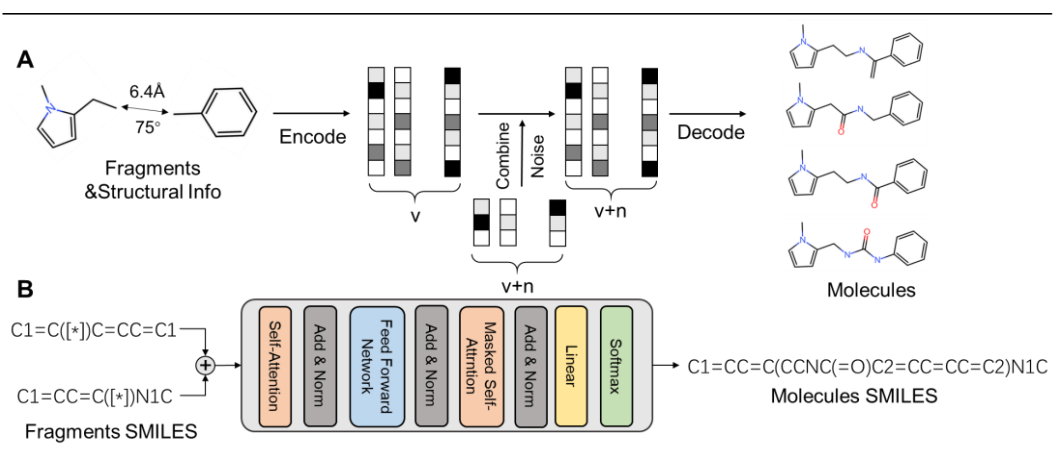

**Figure 5.** Two representative Linker Design methods. A). Delinker illustration B).

SyntaLinker illustration.

While prior deep probabilistic models aimed to broaden the chemical space accessible for linker design by learning and resampling the probability distribution of linkers given fragments p(Linker|Fragments), another storyline lies in Reinforcement Learning (RL). This approach allows researchers to utilize reinforcement learning strategies to generate linker structures by focusing on specific molecular properties like Quantitative Estimate of Drug-likeness[59] (QED), Synthetic Accessibility[60] (SA), and LogP, and even enables a structure-based strategy through the incorporation of protein-ligand interaction energy into the reward functions. Based on this methodology, some researchers have developed several RL-based models to control specific properties of generated linkers. DRLinker[46] is an advanced version of SyntaLinker enhanced by the reinforcement learning strategy, in which the provided input could be, for instance, the linker length, physicochemical properties (QED, SA, LopP) and biological activity of the linker. The success rate of several conditions is all above 90%, except for controlling the length of the linker, proving the effectiveness of reinforcement learning in controlling drug-forming properties. Link-INVENET[61] is an RNN-based method that generates molecules in SMILES format and performs RL on several metrics similar to DRLinker.

Designing effective linkers for lead optimization requires careful consideration of the geometric configuration between the fragments and the chemical environment within the binding pocket. The linker should allow the linked molecule to bind tightly to the target while maintaining the original binding pattern of the lead fragment. In this context, it is important to take into account the detailed geometry-aware interactions within the pocket, as this can not only help bypass pocket hindrances but also further improve the binding affinity of the lead compound by introducing additional energy through linker-pocket interactions. While existing graph-based linker design models, such as DeLinker and DEVELOP, consider the positional and orientational relationships between two fragments by modeling the conditional distribution

$p(G_{Linker}|G_{Frags}, R_{Frags})$, other transformer-based linker design models only complete the linker between two fragment SMILES, modeling the conditional distribution $p(G_{Linker}|G_{Frags})$. As a result, these models are limited to generating two-dimensional linker structures, without explicitly considering the detailed geometry-aware interactions within protein pockets. The currently developed DiffLinker tackles this problem for the first time by combining the popular diffusion-based model[62] and the equivariant graph neural network[63]. Difflinker can be formulated as $p(R_{Linker}, G_{Linker}|R_{Frags}, G_{Frags}, Prot)$, which means it co-generates linker chemical formulas and conformations. In DiffLinker, model first requires a pre-trained model for predicting the number of atoms in linkers. Once the number of linker atoms is determined, the denoising process[64] would be performed on atom type space and interatomic distance space, and the linker with geometric structure is finally generated by evolving the reverse diffusion process from the Gaussian distribution. Another approach, ShapeLinker[65], introduces a dual-phase linker design protocol, making the model aware of protein structures, i.t., $p(G_{Linker}|R_{Frags}, G_{Frags}, Prot)$. Initially, it utilizes RDKit's built-in method[66] to obtain conformations of the generated linker, followed by employing a rapid shape alignment score to assess the compatibility between the generated molecules and protein pockets. While this dual-phase strategy effectively conditions ShapeLinker on protein structures, its reliance on traditional conformer generation methods in a pocket-free context necessitates the generation of numerous potential conformations to ensure compatibility with the protein pocket. This element of randomness could result in the model overlooking molecules that reveal suitable geometric configurations only when aligned with the protein-induced fit, thus potentially missing out on optimal molecular matches. Finally, the LinkerNet[67] model introduces a Newton-Euler-inspired module to modify the initial fragment coordinates, incorporating a kind of flexibility in linker design. LinkerNet can be written as $p(R_{Linker}, G_{Linker}, R_{Frags}|G_{Frags}, Prot)$.

**Scaffold Hopping**

Scaffold hopping is another common strategy used in drug discovery to identify novel compounds with similar pharmacological properties to existing lead molecules but with different chemical scaffolds[10]. This approach can be achieved through computational methods like BM scaffolds or empirical knowledge of medicinal chemists. The goal of scaffold hopping is to optimize the lead scaffold and produce a novel compound with similar properties that can potentially modulate drug-likeness properties while circumventing existing patent protection[68-70]. Traditional methods for scaffold hopping include similarity search, pharmacophore matching and fragment substitution. Despite the diversity in approaches, they all fundamentally address the challenge of comparing similarities between molecules or scaffolds. For a long time, whereas the scaffold hopping has been slowly evolving, there has been no major breakthrough in the underlying methodology, until data-driven representative learning breathed new vigor into it. It is worth noting that the linker itself can be divided into subsets of the scaffold. Therefore, some deep learning-based scaffold hopping models are basically linker design models, such as SyntaLinker, Link-INVNET, DRLinker, etc. These models are limited to linker-like scaffolds between two fragments and cannot be extended to more complex tasks such as those involving multiple fragments or those that occupy a significant portion of the lead compound.

One of the models designed specifically for scaffold hopping is Graph-GMVAE[71], which adopts an MGVAE architecture[72] with a change in the hidden space from the single Gaussian distribution to a multivariate Gaussian distribution. This modification incorporates chemical intuition and enables the definition of similarity between scaffold clusters (each Gaussian area) in the hidden space. Sampling in the hidden space of GMVAE that is given a similarity metric allows for three scales of skeleton hopping: crawling, hopping, and leaping, as proposed in its original paper, shown in **Figure 6A**. Moreover, Graph-GMVAE has two channels, the side-chain channel and the scaffold channel, and the overall generation scheme involves sampling from the scaffold channel interacting with the side-chain channel. ScaffoldGVAE[73] adopts a similar strategy to Graph-GMVAE but releases their code as open-source, providing an easy-to-use model for drug design community.

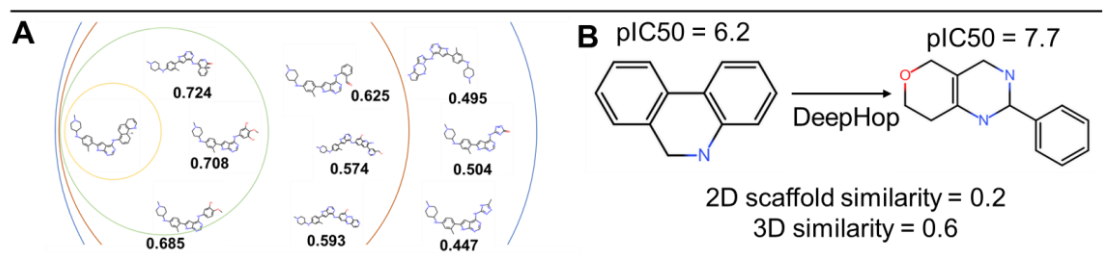

**Figure 6.** Two representative Scaffold Hopping methods. A). GraphGMVAE illustration, different colors represent different degrees of hop B). DeepHop needs to satisfy three criteria.

The DeepHop[23] method, on the other hand, transforms the scaffold hopping into a language translation task. It takes an existing molecule as input and generates a lead molecule with a similar three-dimensional structure but a different two-dimensional structure as output. DeepHop has a special data handling method, which defines a training pair (X,Y), where X is the original molecule and Y is the leap molecule, which satisfies the following three conditions: (1) the scaffold similarity is less than 0.6 (2) the 3D similarity is more than 0.6 (3) the p-bioactivity of Y is one unit higher than that of X. The example of transformation is demonstrated in **Figure 6B**. Based on these pre-processed scaffold leap pairs, DeepHop translates the lead molecule into an optimized molecule, increasing biological activity increasing and changing the topology while maintaining the 3D structure within a certain range. However, DeepHop cannot guarantee the occurrence of a given structure within the hop molecule, which makes it challenging to interact with chemists. Besides, its target-specific capability relies on mapping an entire protein structure to a vector via an embedding layer, which lacks detailed interactions and only demonstrates its generalization on the conserved kinase family. Currently developed DiffHopp[74] achieved detailed-interaction awareness by utilizing a similar architecture to DiffLinker.

**Sidechains Decoration**

Parallel to scaffold hopping, side chains obtained from scaffold decomposition could also be utilized to optimize lead compounds. This process, known as side-chain decoration or scaffold-constraint molecular generation, involves retaining the privileged scaffold with biological activity while modifying side chains to enhance effects. The scaffold often anchors the binding pattern of the molecule in protein pockets through interactions with key residues, which is crucial for maintaining desired interactions during lead optimization[75]. In addition, a series of scaffold-based derivatives are easier to synthesize from an organic standpoint, which greatly reduces the synthesis difficulty in drug development[76]. Anchoring a molecule securely in a protein pocket is akin to securing a ship in the sea, requiring both the stability of the ship and the precise placement of the anchor on the seabed.

The first deep learning-based method in side-chains elaboration is GraphScaffold[77], which generates novel molecules containing the input scaffolds with certainty by sequentially adding atoms and bonds. Its utilization of graph neural network avoids the ambiguity inherent in SMILES-based models regarding atom or bond addition. In this context, the underlying distribution of G can be expressed as $p(G; S)$, where G is the entire molecular graph and S is a scaffold graph acting as a parametric argument. DeepScaffold[23] makes improvements by expanding the scaffold types and including a much wider range of metrics than GraphScaffold. Besides, it could also sample the chemical scaffold structures from generic anonymous scaffolds by modeling $p(G_{chem-scf}|G_{any-scf})$ as shown in **Figure 7A**. Another method, MoLeR[78], strictly constrains molecular generation on given motifs, adding atoms or predefined motifs to intact scaffold with single bonds, which means it can generate arbitrary structures, such as unique rings, even in the absence of pre-split motifs. Combined with parallel training protocol and advanced optimization method (MSO), MoLeR achieves stable and efficient training and inference speed. GNNGAC[79] enables chemists to modify the chemical bonds or atoms of certain intermediates, facilitating better interaction with expert knowledge. These molecular graph-based methods ensure the validity of the generated compounds and naturally guarantee the occurrence of given substructures in generated molecules. In contrast, the geometric graph-based method, 3D-Scaffold[80], takes 3D coordinates of the chosen scaffold as input and generates 3D coordinates of novel therapeutic candidates as output. By utilizing G-SphereNet[81], the geometry generation part in 3D-Scaffold obeys the physical equivariance constraints. 3D-Scaffold-RL[82] steps further by incorporating a graph-based binding probability predictor into the reward function, which evaluates the binding potency of the 3D structure of the protein pocket and the interacting ligand, thereby making the model protein-structure-aware. In 3D-Scaffold-RL[82], it is proven that the inclusion of the binding predictor significantly increases the binding probability of the generated ligands, elevating it from ~0.25 to ~0.75.

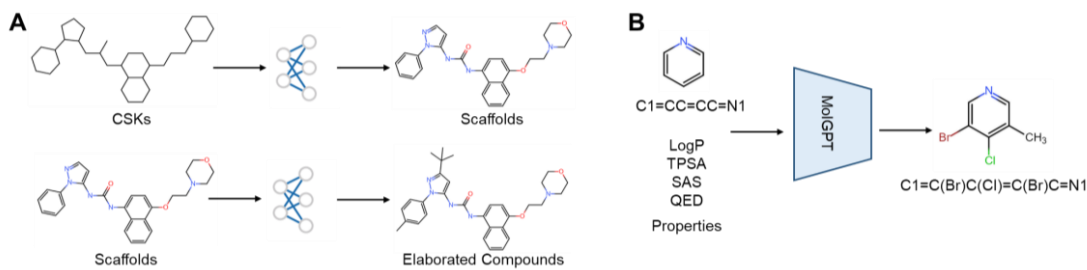

**Figure 7.** Two representative Sidechain Decoration methods. A). DeepScaffold illustration B). MolGPT illustration

Another methodology to optimize the functionality of lead scaffold is from a chemical language perspective, which suffers from illegality when adding fragments to the scaffold. BiDesign[83] alleviates this problem by generating molecules at both ends of SMILES, but it is limited to cases with only two contact points. Scaffold Decorator[84] overcomes the illegality problem by initially defining contact point token * in scaffold. During the generation phase, the language model then grows fragments at these points in a SMILES format. SAMOA devises a new sampling scheme for the SMILES grammar, which makes the scaffold-constraint generation more natural. Not only this sampling scheme could be applied for side-chain decoration, but also it is accessible to the linker design problem. In exception to several models that enforce scaffold constraints in chemical language generation, MolGPT[85] utilizes the latest generative pre-training (GPT) technique[86] in Natural Language Processing (NLP) for molecular conditional generation, which treats scaffold structure as input token and generates a whole completed molecule, as shown in **Figure 7B**. In contrast to the other three SMILES-based side-chain decoration models, MolGPT cannot guarantee that the generated molecules must contain a given substructure, but it can be conditional on several molecular properties like QED and SA by treating them as special tokens. LibINVNET[87] employs reinforcement learning (RL) techniques to achieve similar conditioning and further uses chemical reaction templates to guarantee that the generated molecules are highly synthetic and accessible. LibINVENT can also be aware of protein structures to some extent by integrating docking scores into its reward functions, similar to previous RL-based methods.

Currently developed side-chain decoration models that can both generate molecular 3D structures and be conditional on protein structures is DiffDec[88], an incremental work to the previous DiffLinker and DiffHopp. The development of DiffDec also echoes our insights that one of the important future paths for lead optimization is to venture the structure-based strategy.

**Fragment Replacement**

Fragment replacement, also referred to as fragment growing or substitution, is another important part of fragment-based drug discovery, distinct from the fragment-linking strategy. This optimization task is generally considered more straightforward than linker design since there are fewer geometric constraints, and it has received significant attention[89, 90]. In the early lead discovery stage, fragment growing is applied to complete the lead fragment to fulfill the unoccupied part of the pocket, enhancing the binding strength of the lead molecule. While fragment substitution method bears resemblance to side-chain decoration, they possess distinct characteristics. Under the BM scaffold decomposition, side chains decorated are generally non-cyclic and relatively small, typically offering 4-5 modification sites per compound. However, in fragment replacement, most of the substituted fragments are functional groups containing ring structures. Conventional approaches primarily rely on similarity search, such as

BROOD[91] and FragRep[92]. Some intelligent optimization algorithms have also been used for this task, which is exemplified by AutoGrow[39], which utilizes the genetic algorithm. Deep learning-based fragment replacement methods have not been developed until recent years, such as DeepFrag[22] shown in **Figure 8A**, which reconstructs this problem as a classification problem. In particular, DeepFrag constructs a fragment library and queries the model to determine which fragment should be connected to the seed molecule. Although DeepFrag claims a successful reconstruction rate of more than 50% on the test set, the limitations are obvious. Firstly, the fragment library restricts accessible chemical space, which is also a typical issue for most fragment-based methods. Secondly, it fails to take the geometry of fragment into consideration, which is supposed to be important during the induce-fit process between protein pocket and respective ligands. Furthermore, the architecture of DeepFrag does not satisfy the symmetric property, implying that output would change corresponding to the rotation of input. Thus a brute-force data augment strategy was adopted when training Deepfrag, i.e., randomly rotating the training data while averaging the predictions on 32 uniformly rotated inputs. The subsequently developed DEVELOP[56] method uses the VAE structure to break through the restrictions on fragment libraries. In DEVELOP, pharmacophore information extracted from CNN can also be used to guide the substitution of molecular fragments, as shown in **Figure 8B**. Adopting a similar architecture to DEVELOP, STRIFE[93] uses FHMs[94], a pocket-related descriptor, as a condition for broadening the variety of pharmacophore dependence, making it possible for structure-aware conditional generation. In fragment replacement, a current model that is conditional on detailed protein-ligand interactions is D3FG, which predicts fragments and models the diffusion of geometric distribution in terms of positions and orientations, differing from the atomic position diffusion process in DiffLinker, DiffHopp, and DiffDec. To preserve physical equivariance properties, D3FG designs a novel SO(3) diffusion architecture for orientation predictions.

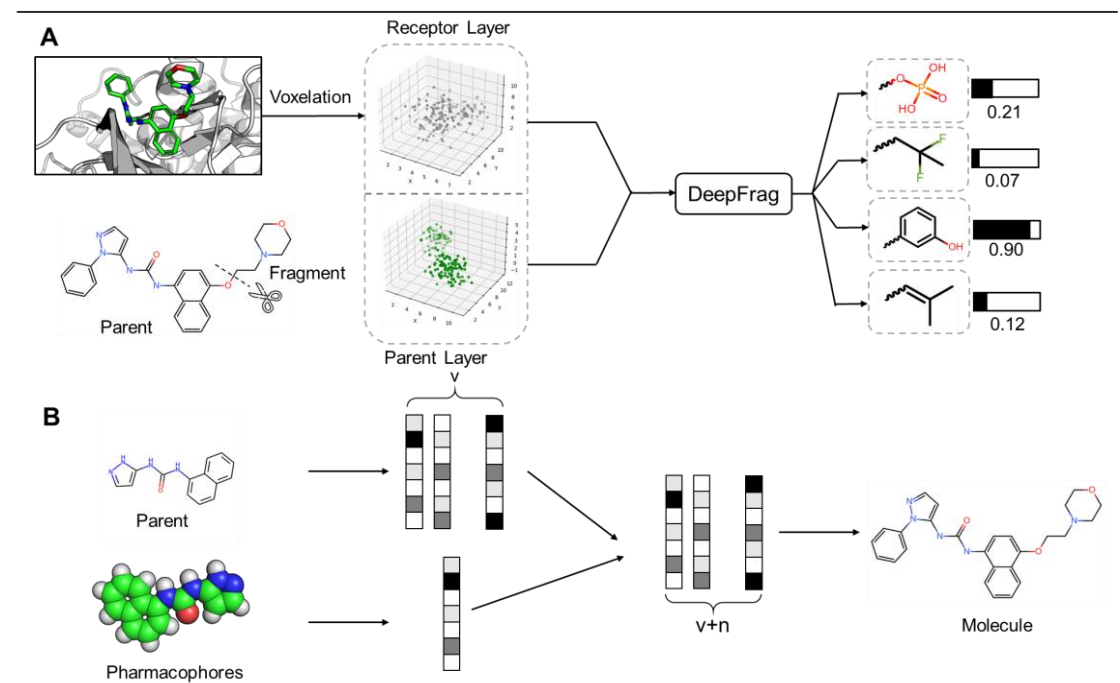

**Figure 8.** Two representative Fragment Growing methods. A). DeepFrag illustration and B). Develop illustration.

**Universal Model**

Through the lens of constrained substructure molecular generation, the lead optimization problem is essentially to complete partial molecules to achieve increment in terms of different properties, like binding affinity, QED, SA, etc. While it might seem trivial to train a singular model to address all four tasks, the reality is more complex. Specifically, inserting fragments at arbitrary points and ensuring connectivity in the SMILES representation poses significant difficulties for SMILES-based generation models, particularly when attempting to complete fragments spatially separated. Many graph-based generation models predefine the growth region, such as DeLinker, which identifies linkers as segments between fragments and utilizes a VAE model to predict linker growth. This approach requires defining different growth modes based on the number of connections to remaining fragments and adapting training accordingly. To address this challenge, four models are proposed, each capable of handling all four lead optimization sub-tasks within a single framework. REINVENT4[95] combines preceding extensions for each task, like LibINVENT and Link-INVENT, into the new platform, creating a comprehensive solution for lead optimization. SAFE[96] introduces a novel representation, redefining SMILES strings as an unordered sequence of interconnected fragment blocks. This representation simplifies substructure-constrained molecular generation to a classical sequence completion task, freeing prior SMILES-based models from designing intricate decoding schemes. Additionally, it adopts the recently popular GPT architecture to train a large model, consistently releasing large-model power in molecular generation. In the graph-based chapter, DrugEXv3[97] introduces a joint graph representation split into three sections: scaffolds, growth, and linking, further detailed by five rows indicating atom type, bond type, connected atom index, atom index, and fragment index. This distinctive setup allows DrugEXv3 to employ a graph transformer model for generating molecules in various substructure-constrained fashions. Despite the unique design of these universal models, they are confined to the 2D ligand generation paradigm. A concurrent model, named Delete[98], steps further to be the first structure-based 3D model of this kind. It employs a unified masking strategy to integrate all tasks within a single model, while also embedding physical equivariance into protein-ligand graph neural networks, thereby enabling the co-design of protein-ligand binding conformations.

## Challenges and Prospects

### Structure-based drug design paradigm

Prediction of drug-target interactions is primarily based on geometric matching and energy matching. Geometric matching depends on the 3D structures of molecules, whereas energy matching is more related to the topology of molecules. These two matching patterns form the basis of structure-based drug design (SBDD) methods[99]. The ultimate goal of SBDD is to create target-specific drug candidates, therefore achieving rational drug design. Nonetheless, many learning-based optimization models are limited to ligand-based paradigms. They can only optimize the 2D structures of molecules with similarly distributed properties to those in the training set. Several approaches that make use of the conditional generative framework only embed proteins to a feature vector and then guide model generation in the hidden space based on this vector. Others utilize reinforcement learning to consider protein target constraints as feedback (e.g., docking score[100]) to generate a series of target-aware molecules. However, the use of reinforcement learning with docking scores is an indirect way of protein-ligand binding interaction information. More direct approaches, like completing 3D fragments within protein pockets, are favored. The emergence of direct pocket-aware 3D lead optimization models like DiffLinker and Delete reflects researchers' increased focus on the advanced structure-based way. Another interesting future direction is to consider protein conformational change, i.e., "Protein flexibility"[101] in structure-based drug design. In the future, it is envisioned that the model development will continue to relax the current rigid assumption. Such flexibility would permit a more physical and rational designing process and increase the likelihood of identifying viable drug candidates.

**Data Construction and Evaluation metrics**

The data required for lead optimization and *de novo* design model differ significantly. The former involves simulating the developmental process from lead compounds to drug candidates, necessitating not only complete molecular structures but also intermediate structures. Unfortunately, such data is scarce, making it insufficient for model training. Nonetheless, it is more compelling to test these models using as much real-world target discovery data as possible, rather than relying solely on a subset of the training data. Thus, there is a need to curate a dataset dedicated to evaluating lead optimization models, which would enable unified comparisons between different models. On the other hand, training data pairs obtained by current substructure decomposition methods (e.g., BRICS, MMP) are still somewhat rudimentary. Therefore, more refined construction methods for training data should be explored, including techniques to introduce slight changes to the original structures during the generation process, such as electron isomer replacement.

Beyond data, most of the current metrics are directly extracted from *de novo* molecular generation models, such as uniqueness, validity, and novelty. Future researchers may consider designing task-dependent metrics, such as assessment of generated scaffold similarity for scaffold hopping, assessment of linker properties for linker design, etc. Although these issues are non-trivial, better metrics will undoubtedly diagnose subtle problems with existing models and, eventually, lead to better models. We anticipate breakthroughs in future models and the development of swift pipelines for experimental validation in real-world drug discovery scenarios.

## Acknowledgments

This work was financially supported by the National Natural Science Foundation of China (22220102001, 92370130, 22303081), China Postdoctoral Science Foundation (2022M722795), and the Baidu Scholarship.

## Reference

1. Torjesen, I., Drug development: the journey of a medicine from lab to shelf. *Pharmaceutical Journal* **2015**.
2. Hevener, K. E.; Pesavento, R.; Ren, J.; Lee, H.; Ratia, K.; Johnson, M. E. Hit-to-Lead: hit validation and assessment. In *Methods in Enzymology*; Elsevier: 2018; Vol. 610, pp 265-309.
3. Joseph-McCarthy, D.; Baber, J. C.; Feyfant, E.; Thompson, D. C.; Humblet, C., Lead optimization via high-throughput molecular docking. *Current opinion in drug discovery & development* **2007**, 10, 264-274.
4. Keserü, G. M.; Makara, G. M., The influence of lead discovery strategies on the properties of drug candidates. *nature reviews Drug Discovery* **2009**, 8, 203-212.
5. Jiang, D.; Hsieh, C.-Y.; Wu, Z.; Kang, Y.; Wang, J.; Wang, E.; Liao, B.; Shen, C.; Xu, L.; Wu, J., Interactiongraphnet: A novel and efficient deep graph representation learning framework for accurate protein–ligand interaction predictions. *Journal of medicinal chemistry* **2021**, 64, 18209-18232.
6. Zang, C.; Wang, F. MoFlow: an invertible flow model for generating molecular graphs. In Proceedings of the 26th ACM SIGKDD International Conference on Knowledge Discovery & Data Mining, 2020; 2020; pp 617-626.
7. Le Guilloux, V.; Schmidtke, P.; Tuffery, P., Fpocket: an open source platform for ligand pocket detection. *BMC bioinformatics* **2009**, 10, 1-11.
8. Zhavoronkov, A.; Ivanenkov, Y. A.; Aliper, A.; Veselov, M. S.; Aladinskiy, V. A.; Aladinskaya, A. V.; Terentiev, V. A.; Polykovskiy, D. A.; Kuznetsov, M. D.; Asadulaev, A., Deep learning enables rapid identification of potent DDR1 kinase inhibitors. *Nature biotechnology* **2019**, 37, 1038-1040.
9. Stokes, J. M.; Yang, K.; Swanson, K.; Jin, W.; Cubillos-Ruiz, A.; Donghia, N. M.; MacNair, C. R.; French, S.; Carfrae, L. A.; Bloom-Ackermann, Z., A deep learning approach to antibiotic discovery. *Cell* **2020**, 180, 688-702. e13.
10. Schneider, G.; Neidhart, W.; Giller, T.; Schmid, G., "Scaffold-hopping" by topological

pharmacophore search: a contribution to virtual screening. *Angewandte Chemie International Edition* **1999**, 38, 2894-2896.

11. Erlanson, D. A.; McDowell, R. S.; O'Brien, T., Fragment-based drug discovery. *Journal of medicinal chemistry* **2004**, 47, 3463-3482.

12. Cross, S.; Cruciani, G., FragExplorer: GRID-based fragment growing and replacement. *Journal of Chemical Information and Modeling* **2022**, 62, 1224-1235.

13. Yao, H.; Liu, J.; Xu, S.; Zhu, Z.; Xu, J., The structural modification of natural products for novel drug discovery. *Expert opinion on drug discovery* **2017**, 12, 121-140.

14. Davidson, M.; McKenney, J.; Stein, E.; Schrott, H.; Bakker-Arkena, R.; Fayyad, R.; Black, D., Comparison of one-year efficacy and safety of atorvastatin versus lovastatin in primary hypercholesterolemia. *American Journal of Cardiology* **1997**, 79, 1475-1481.

15. Becnel Boyd, L.; Maynard, M. J.; Morgan-Linnell, S. K.; Horton, L. B.; Sucgang, R.; Hamill, R. J.; Jimenez, J. R.; Versalovic, J.; Steffen, D.; Zechiedrich, L., Relationships among ciprofloxacin, gatifloxacin, levofloxacin, and norfloxacin MICs for fluoroquinolone-resistant Escherichia coli clinical isolates. *Antimicrobial agents and chemotherapy* **2009**, 53, 229-234.

16. Lim, J.; Ryu, S.; Kim, J. W.; Kim, W. Y., Molecular generative model based on conditional variational autoencoder for de novo molecular design. *Journal of cheminformatics* **2018**, 10, 1-9.

17. Kingma, D. P.; Welling, M., Auto-encoding variational bayes. *arXiv preprint arXiv:1312.6114* **2013**.

18. Dinh, L.; Krueger, D.; Bengio, Y., Nice: Non-linear independent components estimation. *arXiv preprint arXiv:1410.8516* **2014**.

19. Prykhodko, O.; Johansson, S. V.; Kotsias, P.-C.; Arús-Pous, J.; Bjerrum, E. J.; Engkvist, O.; Chen, H., A de novo molecular generation method using latent vector based generative adversarial network. *Journal of Cheminformatics* **2019**, 11, 1-13.

20. Goodfellow, I.; Pouget-Abadie, J.; Mirza, M.; Xu, B.; Warde-Farley, D.; Ozair, S.; Courville, A.; Bengio, Y., Generative adversarial nets. *Advances in neural information processing systems* **2014**, 27.

21. Imrie, F.; Bradley, A. R.; van der Schaar, M.; Deane, C. M., Deep generative models for 3D linker design. *Journal of chemical information and modeling* **2020**, 60, 1983-1995.

22. Green, H.; Koes, D. R.; Durrant, J. D., DeepFrag: a deep convolutional neural network for fragment-based lead optimization. *Chemical Science* **2021**, 12, 8036-8047.

23. Zheng, S.; Lei, Z.; Ai, H.; Chen, H.; Deng, D.; Yang, Y., Deep scaffold hopping with multimodal transformer neural networks. *Journal of cheminformatics* **2021**, 13, 1-15.

24. Wilkens, S. J.; Janes, J.; Su, A. I., HierS: hierarchical scaffold clustering using topological chemical graphs. *Journal of medicinal chemistry* **2005**, 48, 3182-3193.

25. Rombach, R.; Blattmann, A.; Lorenz, D.; Esser, P.; Ommer, B. High-resolution image synthesis with latent diffusion models. In Proceedings of the IEEE/CVF Conference on Computer Vision and Pattern Recognition, 2022; 2022; pp 10684-10695.

26. Goodfellow, I.; Pouget-Abadie, J.; Mirza, M.; Xu, B.; Warde-Farley, D.; Ozair, S.; Courville, A.; Bengio, Y., Generative adversarial networks. *Communications of the ACM* **2020**, 63, 139-144.

27. Lewell, X. Q.; Jones, A. C.; Bruce, C. L.; Harper, G.; Jones, M. M.; Mclay, I. M.; Bradshaw, J., Drug rings database with web interface. A tool for identifying alternative chemical rings in lead discovery programs. *Journal of medicinal chemistry* **2003**, 46, 3257-3274.

28. Shearer, J.; Castro, J. L.; Lawson, A. D.; MacCoss, M.; Taylor, R. D., Rings in clinical trials and

drugs: Present and future. *Journal of Medicinal Chemistry* **2022**, 65, 8699-8712.
29. Bemis, G. W.; Murcko, M. A., The properties of known drugs. 1. Molecular frameworks. *Journal of medicinal chemistry* **1996**, 39, 2887-2893.
30. Koch, M. A.; Schuffenhauer, A.; Scheck, M.; Wetzel, S.; Casaulta, M.; Odermatt, A.; Ertl, P.; Waldmann, H., Charting biologically relevant chemical space: a structural classification of natural products (SCONP). *Proceedings of the National Academy of Sciences* **2005**, 102, 17272-17277.
31. Schuffenhauer, A.; Ertl, P.; Roggo, S.; Wetzel, S.; Koch, M. A.; Waldmann, H., The scaffold tree − visualization of the scaffold universe by hierarchical scaffold classification. *Journal of chemical information and modeling* **2007**, 47, 47-58.
32. Varin, T.; Gubler, H.; Parker, C. N.; Zhang, J.-H.; Raman, P.; Ertl, P.; Schuffenhauer, A., Compound set enrichment: a novel approach to analysis of primary HTS data. *Journal of chemical information and modeling* **2010**, 50, 2067-2078.
33. Varin, T.; Schuffenhauer, A.; Ertl, P.; Renner, S., Mining for bioactive scaffolds with scaffold networks: improved compound set enrichment from primary screening data. *Journal of chemical information and modeling* **2011**, 51, 1528-1538.
34. Wetzel, S.; Klein, K.; Renner, S.; Rauh, D.; Oprea, T. I.; Mutzel, P.; Waldmann, H., Interactive exploration of chemical space with Scaffold Hunter. *Nature chemical biology* **2009**, 5, 581-583.
35. Terrett, N. K., Combinatorial chemistry. **1998**.
36. Dolle, R. E., Comprehensive survey of combinatorial library synthesis: 1999. *Journal of Combinatorial Chemistry* **2000**, 2, 383-433.
37. Naderi, M.; Alvin, C.; Ding, Y.; Mukhopadhyay, S.; Brylinski, M., A graph-based approach to construct target-focused libraries for virtual screening. *Journal of cheminformatics* **2016**, 8, 1-16.
38. Thompson, D. C.; Aldrin Denny, R.; Nilakantan, R.; Humblet, C.; Joseph-McCarthy, D.; Feyfant, E., CONFIRM: connecting fragments found in receptor molecules. *Journal of Computer-Aided Molecular Design* **2008**, 22, 761-772.
39. Durrant, J. D.; Amaro, R. E.; McCammon, J. A., AutoGrow: a novel algorithm for protein inhibitor design. *Chemical biology & drug design* **2009**, 73, 168-178.
40. Böhm, H.-J., LUDI: rule-based automatic design of new substituents for enzyme inhibitor leads. *Journal of computer-aided molecular design* **1992**, 6, 593-606.
41. Lewell, X. Q.; Judd, D. B.; Watson, S. P.; Hann, M. M., Recap retrosynthetic combinatorial analysis procedure: a powerful new technique for identifying privileged molecular fragments with useful applications in combinatorial chemistry. *Journal of chemical information and computer sciences* **1998**, 38, 511-522.
42. Degen, J.; Wegscheid-Gerlach, C.; Zaliani, A.; Rarey, M., On the Art of Compiling and Using'Drug-Like'Chemical Fragment Spaces. *ChemMedChem: Chemistry Enabling Drug Discovery* **2008**, 3, 1503-1507.
43. Liu, T.; Naderi, M.; Alvin, C.; Mukhopadhyay, S.; Brylinski, M., Break down in order to build up: decomposing small molecules for fragment-based drug design with e molfrag. *Journal of chemical information and modeling* **2017**, 57, 627-631.
44. Hussain, J.; Rea, C., Computationally efficient algorithm to identify matched molecular pairs (MMPs) in large data sets. *Journal of chemical information and modeling* **2010**, 50, 339-348.
45. Igashov, I.; Stärk, H.; Vignac, C.; Satorras, V. G.; Frossard, P.; Welling, M.; Bronstein, M.; Correia, B., Equivariant 3d-conditional diffusion models for molecular linker design. *arXiv*

*preprint arXiv:2210.05274* **2022**.

46. Tan, Y.; Dai, L.; Huang, W.; Guo, Y.; Zheng, S.; Lei, J.; Chen, H.; Yang, Y., DRlinker: Deep Reinforcement Learning for Optimization in Fragment Linking Design. *Journal of Chemical Information and Modeling* **2022**, 62, 5907-5917.

47. Jhoti, H.; Williams, G.; Rees, D. C.; Murray, C. W., The'rule of three'for fragment-based drug discovery: where are we now? *Nature Reviews Drug Discovery* **2013**, 12, 644-644.

48. Erlanson, D. A.; Fesik, S. W.; Hubbard, R. E.; Jahnke, W.; Jhoti, H., Twenty years on: the impact of fragments on drug discovery. *Nature reviews Drug discovery* **2016**, 15, 605-619.

49. Iegre, J.; Brear, P.; De Fusco, C.; Yoshida, M.; Mitchell, S. L.; Rossmann, M.; Carro, L.; Sore, H. F.; Hyvönen, M.; Spring, D. R., Second-generation CK2α inhibitors targeting the αD pocket. *Chemical Science* **2018**, 9, 3041-3049.

50. De Fusco, C.; Brear, P.; Iegre, J.; Georgiou, K. H.; Sore, H. F.; Hyvönen, M.; Spring, D. R., A fragment-based approach leading to the discovery of a novel binding site and the selective CK2 inhibitor CAM4066. *Bioorganic & medicinal chemistry* **2017**, 25, 3471-3482.

51. Yu, H. S.; Modugula, K.; Ichihara, O.; Kramschuster, K.; Keng, S.; Abel, R.; Wang, L., General theory of fragment linking in molecular design: why fragment linking rarely succeeds and how to improve outcomes. *Journal of Chemical Theory and Computation* **2020**, 17, 450-462.

52. Ichihara, O.; Barker, J.; Law, R. J.; Whittaker, M., Compound design by fragment-linking. *Molecular Informatics* **2011**, 30, 298-306.

53. Yang, Y.; Zheng, S.; Su, S.; Zhao, C.; Xu, J.; Chen, H., SyntaLinker: automatic fragment linking with deep conditional transformer neural networks. *Chemical science* **2020**, 11, 8312-8322.

54. Vaswani, A.; Shazeer, N.; Parmar, N.; Uszkoreit, J.; Jones, L.; Gomez, A. N.; Kaiser, Ł.; Polosukhin, I., Attention is all you need. *Advances in neural information processing systems* **2017**, 30.

55. Feng, Y.; Yang, Y.; Deng, W.; Chen, H.; Ran, T., SyntaLinker-Hybrid: A deep learning approach for target specific drug design. *Artificial Intelligence in the Life Sciences* **2022**, 2, 100035.

56. Imrie, F.; Hadfield, T. E.; Bradley, A. R.; Deane, C. M., Deep generative design with 3D pharmacophoric constraints. *Chemical science* **2021**, 12, 14577-14589.

57. Gu, J.; Wang, Z.; Kuen, J.; Ma, L.; Shahroudy, A.; Shuai, B.; Liu, T.; Wang, X.; Wang, G.; Cai, J., Recent advances in convolutional neural networks. *Pattern recognition* **2018**, 77, 354-377.

58. Huang, Y.; Peng, X.; Ma, J.; Zhang, M., 3dlinker: An e (3) equivariant variational autoencoder for molecular linker design. *arXiv preprint arXiv:2205.07309* **2022**.

59. Clark, D. E.; Pickett, S. D., Computational methods for the prediction of 'drug-likeness'. *Drug discovery today* **2000**, 5, 49-58.

60. Ertl, P.; Schuffenhauer, A., Estimation of synthetic accessibility score of drug-like molecules based on molecular complexity and fragment contributions. *Journal of cheminformatics* **2009**, 1, 1-11.

61. Guo, J.; Knuth, F.; Margreitter, C.; Janet, J. P.; Papadopoulos, K.; Engkvist, O.; Patronov, A., Link-INVENT: Generative Linker Design with Reinforcement Learning. **2022**.

62. Song, J.; Meng, C.; Ermon, S., Denoising diffusion implicit models. *arXiv preprint arXiv:2010.02502* **2020**.

63. Satorras, V. G.; Hoogeboom, E.; Fuchs, F. B.; Posner, I.; Welling, M., E (n) equivariant normalizing flows. *arXiv preprint arXiv:2105.09016* **2021**.

64. Ho, J.; Jain, A.; Abbeel, P., Denoising diffusion probabilistic models. *Advances in neural information processing systems* **2020**, 33, 6840-6851.

65. Neeser, R. M.; Akdel, M.; Kovtun, D.; Naef, L., Reinforcement Learning-Driven Linker Design via Fast Attention-based Point Cloud Alignment. *arXiv preprint arXiv:2306.08166* **2023**.

66. Crippen, G. M.; Havel, T. F., *Distance geometry and molecular conformation*. Research Studies Press Taunton: 1988; Vol. 74.

67. Guan, J.; Peng, X.; Jiang, P.; Luo, Y.; Peng, J.; Ma, J., LinkerNet: Fragment Poses and Linker Co-Design with 3D Equivariant Diffusion. *Advances in Neural Information Processing Systems* **2024**, 36.

68. Cramer, R. D.; Jilek, R. J.; Guessregen, S.; Clark, S. J.; Wendt, B.; Clark, R. D., "Lead hopping". Validation of topomer similarity as a superior predictor of similar biological activities. *Journal of medicinal chemistry* **2004**, 47, 6777-6791.

69. Brown, N.; Jacoby, E., On scaffolds and hopping in medicinal chemistry. *Mini reviews in medicinal chemistry* **2006**, 6, 1217-1229.

70. Martin, Y. C.; Muchmore, S., Beyond QSAR: lead hopping to different structures. *QSAR & Combinatorial Science* **2009**, 28, 797-801.

71. Yu, Y.; Xu, T.; Li, J.; Qiu, Y.; Rong, Y.; Gong, Z.; Cheng, X.; Dong, L.; Liu, W.; Li, J., A novel scalarized scaffold hopping algorithm with graph-based variational autoencoder for discovery of JAK1 inhibitors. *ACS omega* **2021**, 6, 22945-22954.

72. Lee, D. B.; Min, D.; Lee, S.; Hwang, S. J. Meta-gmvae: Mixture of gaussian vae for unsupervised meta-learning. In International Conference on Learning Representations, 2021; 2021.

73. Hu, C.; Li, S.; Yang, C.; Chen, J.; Xiong, Y.; Fan, G.; Liu, H.; Hong, L., ScaffoldGVAE: scaffold generation and hopping of drug molecules via a variational autoencoder based on multi-view graph neural networks. *Journal of Cheminformatics* **2023**, 15, 91.

74. Torge, J.; Harris, C.; Mathis, S. V.; Lio, P., Diffhopp: A graph diffusion model for novel drug design via scaffold hopping. *arXiv preprint arXiv:2308.07416* **2023**.

75. Welsch, M. E.; Snyder, S. A.; Stockwell, B. R., Privileged scaffolds for library design and drug discovery. *Current opinion in chemical biology* **2010**, 14, 347-361.

76. Feng, M.; Tang, B.; H Liang, S.; Jiang, X., Sulfur containing scaffolds in drugs: synthesis and application in medicinal chemistry. *Current topics in medicinal chemistry* **2016**, 16, 1200-1216.

77. Lim, J.; Hwang, S.-Y.; Moon, S.; Kim, S.; Kim, W. Y., Scaffold-based molecular design with a graph generative model. *Chemical science* **2020**, 11, 1153-1164.

78. Maziarz, K.; Jackson-Flux, H.; Cameron, P.; Sirockin, F.; Schneider, N.; Stiefl, N.; Segler, M.; Brockschmidt, M., Learning to extend molecular scaffolds with structural motifs. *arXiv preprint arXiv:2103.03864* **2021**.

79. Hu, S.; Takigawa, I.; Xiao, C., Edit-Aware Generative Molecular Graph Autocompletion for Scaffold Input. **2022**.

80. Joshi, R. P.; Gebauer, N. W.; Bontha, M.; Khazaieli, M.; James, R. M.; Brown, J. B.; Kumar, N., 3D-Scaffold: A deep learning framework to generate 3d coordinates of drug-like molecules with desired scaffolds. *The Journal of Physical Chemistry B* **2021**, 125, 12166-12176.

81. Luo, Y.; Ji, S. An autoregressive flow model for 3d molecular geometry generation from scratch. In International Conference on Learning Representations, 2021; 2021.

82. McNaughton, A. D.; Bontha, M. S.; Knutson, C. R.; Pope, J. A.; Kumar, N., De novo design of

protein target specific scaffold-based Inhibitors via Reinforcement Learning. *arXiv preprint arXiv:2205.10473* **2022**.

83. Grisoni, F.; Moret, M.; Lingwood, R.; Schneider, G., Bidirectional molecule generation with recurrent neural networks. *Journal of chemical information and modeling* **2020**, 60, 1175-1183.

84. Arús-Pous, J.; Patronov, A.; Bjerrum, E. J.; Tyrchan, C.; Reymond, J.-L.; Chen, H.; Engkvist, O., SMILES-based deep generative scaffold decorator for de-novo drug design. *Journal of cheminformatics* **2020**, 12, 1-18.

85. Bagal, V.; Aggarwal, R.; Vinod, P.; Priyakumar, U. D., MolGPT: molecular generation using a transformer-decoder model. *Journal of Chemical Information and Modeling* **2021**, 62, 2064-2076.

86. Radford, A.; Narasimhan, K.; Salimans, T.; Sutskever, I., Improving language understanding by generative pre-training. **2018**.

87. Fialková, V.; Zhao, J.; Papadopoulos, K.; Engkvist, O.; Bjerrum, E. J.; Kogej, T.; Patronov, A., LibINVENT: reaction-based generative scaffold decoration for in silico library design. *Journal of Chemical Information and Modeling* **2021**, 62, 2046-2063.

88. Xie, J.; Chen, S.; Lei, J.; Yang, Y., DiffDec: Structure-Aware Scaffold Decoration with an End-to-End Diffusion Model. *Journal of Chemical Information and Modeling* **2024**.

89. Chevillard, F.; Kolb, P., SCUBIDOO: a large yet screenable and easily searchable database of computationally created chemical compounds optimized toward high likelihood of synthetic tractability. *Journal of chemical information and modeling* **2015**, 55, 1824-1835.

90. Sommer, K.; Flachsenberg, F.; Rarey, M., NAOMInext–Synthetically feasible fragment growing in a structure-based design context. *European Journal of Medicinal Chemistry* **2019**, 163, 747-762.

91. Wang, L.-h.; Evers, A.; Monecke, P.; Naumann, T., Ligand based lead generation–considering chemical accessibility in rescaffolding approaches via BROOD. *Journal of Cheminformatics* **2012**, 4, 1-1.

92. Shan, J.; Pan, X.; Wang, X.; Xiao, X.; Ji, C., FragRep: A Web Server for Structure-Based Drug Design by Fragment Replacement. *Journal of Chemical Information and Modeling* **2020**, 60, 5900-5906.

93. Hadfield, T. E.; Imrie, F.; Merritt, A.; Birchall, K.; Deane, C. M., Incorporating target-specific pharmacophoric information into deep generative models for fragment elaboration. *Journal of Chemical Information and Modeling* **2022**, 62, 2280-2292.

94. Radoux, C. J.; Olsson, T. S.; Pitt, W. R.; Groom, C. R.; Blundell, T. L., Identifying interactions that determine fragment binding at protein hotspots. *Journal of medicinal chemistry* **2016**, 59, 4314-4325.

95. Loeffler, H. H.; He, J.; Tibo, A.; Janet, J. P.; Voronov, A.; Mervin, L. H.; Engkvist, O., Reinvent 4: Modern AI–driven generative molecule design. *Journal of Cheminformatics* **2024**, 16, 20.

96. Noutahi, E.; Gabellini, C.; Craig, M.; Lim, J. S.; Tossou, P., Gotta be SAFE: A New Framework for Molecular Design. *Digital Discovery* **2024**.

97. Liu, X.; Ye, K.; van Vlijmen, H. W.; IJzerman, A. P.; van Westen, G. J., DrugEx v3: scaffold-constrained drug design with graph transformer-based reinforcement learning. *Journal of Cheminformatics* **2023**, 15, 24.

98. Zhang, H.; Zhao, H.; Zhang, X.; Su, Q.; Du, H.; Shen, C.; Wang, Z.; Li, D.; Pan, P.; Chen, G., Delete: Deep Lead Optimization Enveloped in Protein Pocket through Unified Deleting Strategies and a Structure-aware Network. *arXiv preprint arXiv:2308.02172* **2023**.

99. Ebalunode, J. O.; Ouyang, Z.; Liang, J.; Zheng, W., Novel approach to structure-based pharmacophore search using computational geometry and shape matching techniques. *Journal of chemical information and modeling* **2008**, 48, 889-901.

100. Trott, O.; Olson, A. J., AutoDock Vina: improving the speed and accuracy of docking with a new scoring function, efficient optimization, and multithreading. *Journal of computational chemistry* **2010**, 31, 455-461.

101. Jain, A. N., Surflex: fully automatic flexible molecular docking using a molecular similarity-based search engine. *Journal of medicinal chemistry* **2003**, 46, 499-511.